\documentclass[article,twocolumn,english,secnumarabic,footinbib,tightenlines,nobibnotes,aps,prl,unsortedaddress,showpacs,superscriptaddress]{revtex4-1}
\usepackage{graphicx}
\usepackage[range-units=single,separate-uncertainty]{siunitx}
\usepackage[color links=true,linkcolor=blue,citecolor=blue,urlcolor=blue]{hyperref}

\usepackage[usenames,dvipsnames]{color}
\usepackage{amsmath}
\usepackage{amsfonts}
\usepackage{amssymb}
\usepackage{mathtools}
\usepackage{pict2e}
\usepackage{hyperref}
\usepackage{graphicx,xcolor}
\usepackage{bbding}
\usepackage{bm}
\usepackage{bbm}

\usepackage{multirow}
\usepackage{dcolumn}  
\usepackage{float}
\usepackage{amsmath,mathrsfs}
\usepackage{amsthm}
\usepackage{epsfig}
\usepackage{array}
\usepackage{color}
\usepackage{soul}
\usepackage[caption=false]{subfig}

\usepackage{comment}

\newcolumntype{P}[1]{>{\centering\arraybackslash}p{#1}}

\def\blue{\textcolor{blue}}

\def\ie{{\it i.e.\/}}

\mathchardef\sOmega="710A
\mathchardef\sGamma="7100
\mathchardef\sDelta="7101

\def\frac#1#2{{\textstyle{#1 \over #2}}}

\DeclareMathOperator{\sign}{sign}
\newcommand{\nvec}[1]{\textbf{#1}}

\newcommand{\im}{i}
\newcommand{\ltspice}{\textit{LTspice }}
\newcommand{\ii}{\im}
\newcommand{\dd}{d}
\newcommand{\unity}{\mathbbm{1}}
\newcommand{\e}{e}

\begin{document}

\title{Chiral voltage propagation and calibration in a topolectrical Chern circuit}

\author{Tobias Hofmann} 
\thanks{Both authors equally contributed to this work.}
\affiliation{Institute for Theoretical Physics and Astrophysics, University of W\"urzburg, D-97074 W\"urzburg, Germany}
\author{Tobias Helbig} 
\thanks{Both authors equally contributed to this work.}
\affiliation{Institute for Theoretical Physics and Astrophysics, University of W\"urzburg, D-97074 W\"urzburg, Germany}
\author{Ching Hua Lee}
\affiliation{Department of Physics, National University of Singapore, Singapore, 117542.}
\affiliation{Institute of High Performance Computing, A*STAR, Singapore, 138632.}
\author{Martin Greiter} 
\affiliation{Institute for Theoretical Physics and Astrophysics, University of W\"urzburg, D-97074 W\"urzburg, Germany}
\author{Ronny Thomale}
\email{Corresponding author: rthomale@physik.uni-wuerzburg.de}
\affiliation{Institute for Theoretical Physics and Astrophysics, University of W\"urzburg, D-97074 W\"urzburg, Germany}

\date{\today}


\begin{abstract}
We propose an electric circuit array with topologically protected
uni-directional voltage modes at its boundary. Instead of external
bias fields or floquet engineering, we employ negative impedance
converters with current inversion (INICs) to accomplish a
non-reciprocal, time-reversal symmetry broken electronic network we
call topolectrical Chern circuit (TCC). The TCC features an admittance
bulk gap fully tunable via the resistors used in the INICs, along with
a chiral voltage boundary mode reminiscent of the Berry flux monopole
present in the admittance band structure. The active circuit elements
in the TCC can be calibrated to compensate for dissipative loss.  
\end{abstract}
\maketitle


{\it Introduction.} The Chern insulator is the mother state of
topological band theory. Originally conceived by Haldane as a
tight-binding model of electrons with broken time-reversal symmetry on
a hexagonal lattice~\cite{PhysRevLett.61.2015}, it roots in the Berry
phase experienced by the electrons as the Brillouin zone is viewed as a
compact parameter space~\cite{Berry45,PhysRevLett.62.2747}. The lattice Chern number
$C$ is quantized to take integer values, as it counts the total charge
of Berry flux monopoles. For a Chern insulator with open boundaries,
this implies $C$ chiral edge modes which experience topological
protection against any kind of disorder and other imperfections, as
there is no backscattering. This induces a stronger
protection than, for instance, topological insulators, where only
elastic backscattering is prohibited by the symmetry-protected
topological character. Still, dissipative loss is a severe
limitation for Chern insulators, and constitutes the central challenge
to realize stable edge mode propagation.

As the Berry phase is a phenomenon of parameter space and does not
rely on any property of the phase space of quantum electrons, the
Chern insulator suggests itself for a plethora of alternative realizations.
Haldane and Raghu employed this
insight to propose a Chern insulator in photonic crystals by use of
the Faraday effect, where chiral edge modes would manifest as one-way
waveguides~\cite{PhysRevLett.100.013904}. This work inspired the
subsequent formulation and realization of Chern bands in
magneto-optical photonic
crystals~\cite{PhysRevLett.100.013905,marin1}, optical waveguides
subject to a magnetic field~\cite{PhysRevLett.100.023902} or floquet
modulation~\cite{alex1}, ultra-cold atomic gases~\cite{jotzu},
mechanical gyrotropic~\cite{kalu,nash,PhysRevLett.115.104302} and
acoustic~\cite{PhysRevLett.114.114301,alu1} systems, as well as, most
recently, coupled optical resonators~\cite{Bandreseaar4005} and
exciton polariton metamaterials~\cite{klembt}. The nature and
potential technological use of topological chiral edge modes crucially
depends on the constituent degrees of freedom, the magnitude of the bulk gap, 
and the ability to prevent loss from affecting the edge
dynamics. In all beforementioned physical systems, the
latter is the most challenging aspect since, unless
one intends to pump the Chern mode anyway, the edge signal exhibits
significant decay despite its topological protection.

In this Article, we propose a Chern circuit 
which is formed by the
admittance band structure of an electric network. As initially
accomplished for the circuit analogue of a topological crystalline
insulator~\cite{PhysRevLett.106.106802,PhysRevX.5.021031,PhysRevLett.114.173902},
topolectrical circuits~\cite{lee1,luling} have recently been found to
host topological admittance band structures~\cite{helbig1} of high
complexity, including Weyl bands~\cite{lee1,luo,simon1} as well as
higher-order topological insulators~\cite{imhof1,sebi1}. Moving beyond the realm of RLC circuits, the combined
time reversal symmetry and circuit reciprocity breaking through
negative impedance converters with current inversion (INICs)~\cite{Chen} allow us
to formulate a topolectrical Chern circuit (TCC) without external bias
fields or floquet engineering. We find topologically protected chiral
voltage edge modes which, from the viewpoint of electrical
engineering, bear resemblance to a voltage circulator.  In contrast
to previous Chern band realizations, our
arrangement of active circuit elements allows for a
recalibration of gain and loss to protect the topological chiral voltage
signal from decay.


{\it Topolectrical Chern circuit.}
\begin{figure*}[t]
    \centering
    \includegraphics[width=\linewidth]{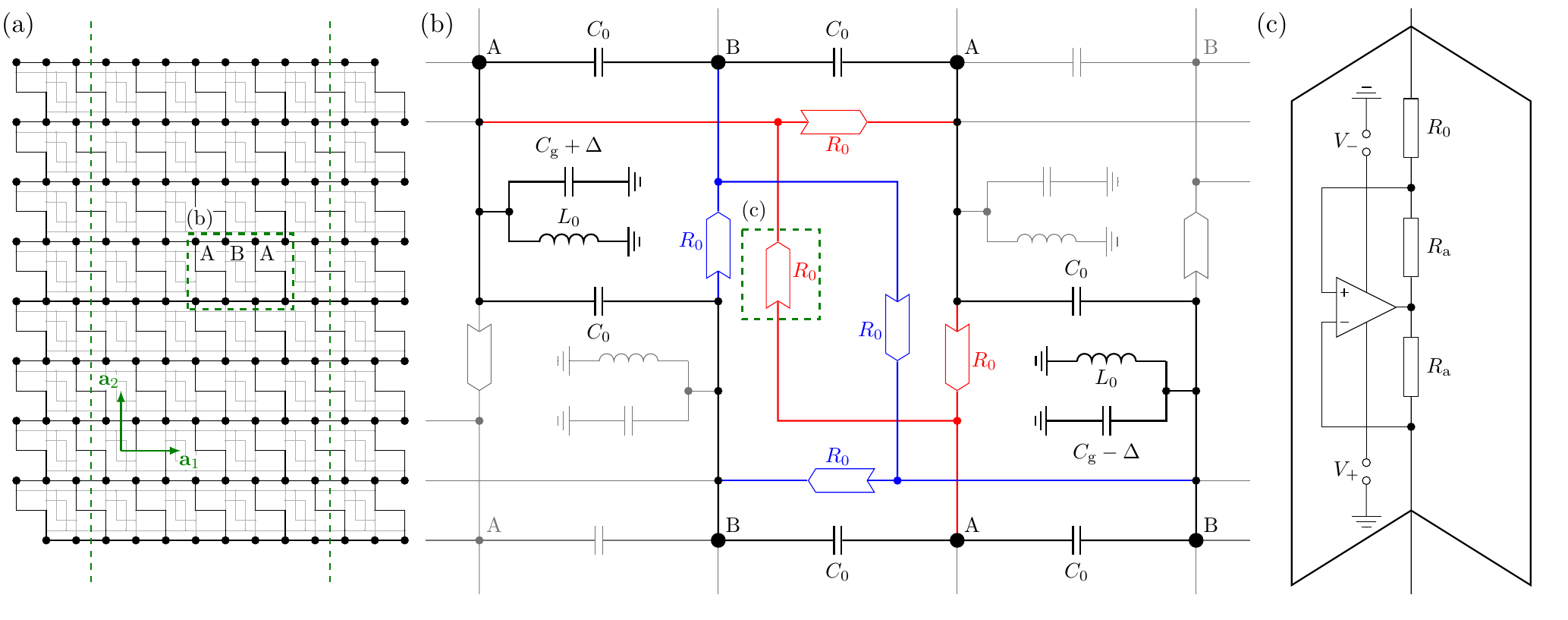}
    \caption{Topolectrical Chern circuit. (a) The three-coordinated
      circuit graph in a brick wall representation of horizontal ($x$)
      and vertical ($y$) alignment of nodes, where the circuit unit
      cell is given by
      two "sublattice'' nodes A and B, the Bravais vectors by
      ${\bf a}_1$ and ${\bf a}_2$, and the $x$ terminations by the
      vertical green dashed lines. (b) The circuit element structure
      is detailed for the green dashed framed
      rectangle in (a). Aside from capacitive inter-node connections
      $C_0$, there are inductive ($L_0$) and capacitive
      ($C_{\text{g}}\pm \Delta$) connections to
      ground. Further A-A and B-B circuit elements are two oppositely
      circular sets of INICs (blue and red) labelled by their
      resistive parameter $R_0$. (c) The INIC element
      structure is shown for the green dashed framed rectangle in
      (b). The arrangement of resistors $R_a$ and $R_0$ combined with an
      operational amplifier with supply voltages $V_+$ and $V_-$ acts
      as a negative impedance converter with current inversion, i.e., as a positive
      (negative) resistor from the front (back) end~\cite{supp}.}
    \label{fig:circuit_diagram}
\end{figure*}
The TCC is formed by a periodic circuit structure sketched
in Fig.~\ref{fig:circuit_diagram}a. The circuit unit cell detailed in
Fig.~\ref{fig:circuit_diagram}b consists of two nodes each of which is
connected to three adjacent nodes through a capacitor $C_0$ and to six
next-nearest neighbours through INICs~\cite{supp}, which are further characterized
in Fig.~\ref{fig:circuit_diagram}c. As discussed in detail below,
these INICs provide the circuit non-reciprocity necessary to induce chiral propagation. The nodes are
grounded by inductors $L_0$ as well as capacitors of capacitance
$C_\text{g} \pm \Delta$ on alternating sublattices $A$ and $B$. 
Due to the graph nature of electric circuits implying a gauge
degree of freedom for arranging the circuit components in real space~\cite{helbig1},
we fix the Bravais vectors as $\nvec{a}_1 = (1,0)$ and $\nvec{a}_2 = (0,1)$ amounting to a
brick wall structure shown in Fig.~\ref{fig:circuit_diagram}a. The
grounded circuit Laplacian $J$ is defined as the matrix relating the vector of voltages $\nvec{V}$ measured with respect to ground 
to the vector of input currents $\nvec{I}$ at the $N$ circuit nodes by $\nvec{I} = J \nvec{V}$. For
an AC frequency $\omega = 2\pi \, f$ and two-dimensional reciprocal space
implied by the brick wall gauge, the TCC Laplacian~\cite{lee1}
$J_{\text{TCC}}(\nvec{k}; \omega)$ and its corresponding spectrum of
eigenvalues $j_{\text{TCC}}(\nvec{k};\omega)$ reads
\begin{widetext}
\begin{align}
J_{\text{TCC}}(\nvec{k}; \omega) &= \ii \omega \bigg[ \left(3 C_0 + C_\text{g} - \frac{1}{\omega^2 L_0}\right) \unity 
- C_0 \left(1 + \cos(k_x) + \cos(k_y)\right) \sigma_x 
- C_0 \left(\sin(k_x) + \sin(k_y)\right) \sigma_y \nonumber \\
&\qquad +\left[ \Delta + \frac{2}{\omega R_0} \left( \sin(k_x) - \sin(k_y) - \sin(k_x - k_y) \right) \right] \sigma_z \bigg], \label{eq:TCC Laplacian} \\
j_{\text{TCC}}(\nvec{k};\omega) &= \ii \omega \bigg[ \left( 3 C_0 + C_\text{g} -\frac{1}{\omega^2} L_0 \right) \pm \sqrt{C_0^2 \left(3+2 \cos(k_x)+2 \cos(k_x-k_y) +2 \cos(k_y)\right)+ (\Delta + \frac{\Gamma(\nvec{k})}{\omega } )^2}\bigg],\label{eq:band structure}
\end{align}
\end{widetext}
where $\Gamma(\nvec{k}) =\frac{2}{R_0} \left(\sin(k_x) - \sin(k_y) -
\sin(k_x - k_y)\right)$. In Fig.~\ref{fig:band structure}a, we show
the projected band structure $j_{\text{TCC;x}}(k_y)$ employing
open-boundary conditions in $x$-direction, as specified in
Fig.~\ref{fig:circuit_diagram}a. It features edge-localized
chiral admittance modes residing in the admittance gap, reminiscent of
the chiral energy modes of a Chern insulator. 
Spectral reflection symmetry of $j_{\text{TCC;x}}(k_y)$ around zero
admittance (Fig.~\ref{fig:band structure}a) is accomplished
for the frequency $\omega_0= 2\pi \, f_0 = 1/\sqrt{3C_0L_0}$.
 
 \begin{figure*}[t]
	\centering
	\includegraphics[width=\linewidth]{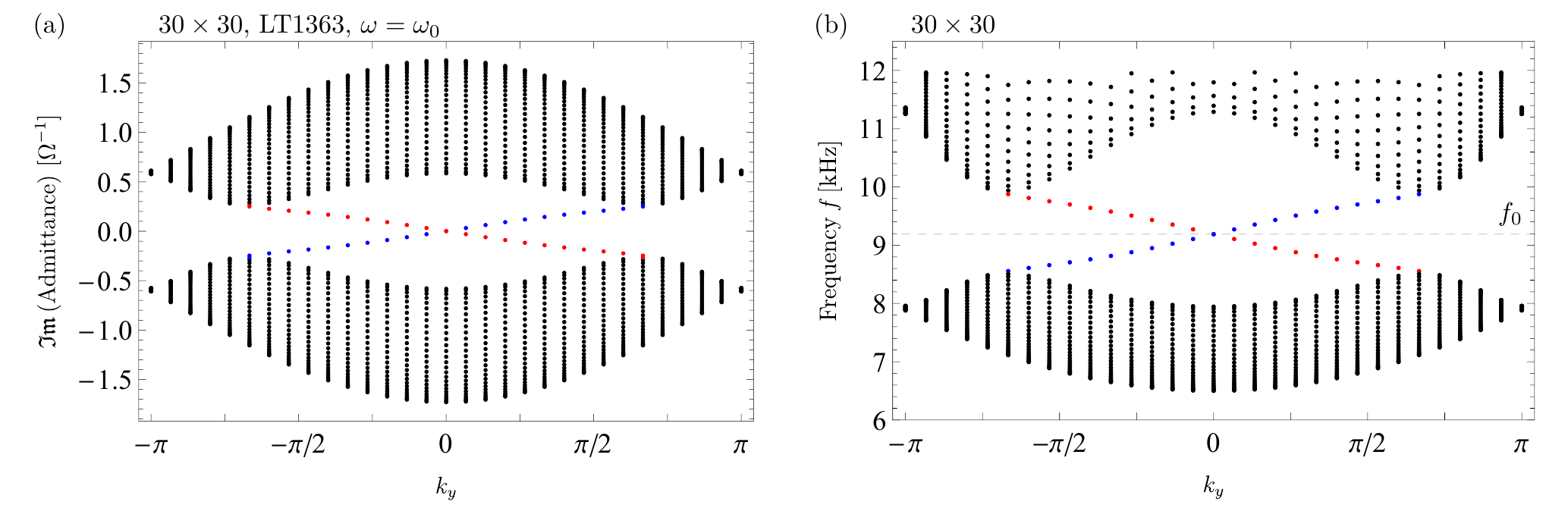}
	\caption{TCC band structure with a boundary.
(a) Admittance band structure $j_{\text{TCC};x}(k_y)$
		obtained from \ltspice simulations for open boundary
		conditions in $x$ direction with a B-A termination (Fig.~\ref{fig:circuit_diagram}a) and periodic boundary conditions
		in $y$-direction. The system size is $30 \times 30 $ unit cells. (b) Frequency band structure
		$\omega(k_y)$ obtained from a numerical calculation for the
		same setting as (a) using the Hamiltonian formalism. TCC parameters are $C_0 = \SI{10}{\micro \farad}$, $L_0 = \SI{10}{\micro \henry}$, $R_0 = \SI{20}{\ohm}$, $C_\text{g}=\SI{0}{\farad}$, $\Delta= \SI{0}{\farad}$ and $f= \SI{9.188}{\kilo \hertz}$. The operational amplifiers are implemented by the \ltspice model LT1363. A chiral boundary mode for the left
		and right $x$ termination (blue and red) is seen in
		the bulk admittance and bulk frequency gap.}
	\label{fig:band structure}
\end{figure*}

{\it Chiral edge modes.} 
The capacitive grounding parameter $\Delta$ appears in \eqref{eq:TCC Laplacian} as an inversion symmetry breaking term reminiscent
of a Semenoff mass~\cite{PhysRevLett.53.2449}. The topological character of the model roots in
the INIC next-nearest neighbour A-A and B-B coupling elements. They break
both time-reversal symmetry and circuit reciprocity via the effective implementation of a negative and positive resistance in the forward and reversed direction of the element, respectively~\cite{supp}.
INICs with clockwise
orientation (Fig. \ref{fig:circuit_diagram}b) effectively act as a voltage circulator, where a voltage profile can only travel in one direction~\cite{supp}. 
This implements a circular motion of voltage in the bulk of the TCC
with no effective translational propagation in any direction. 
The INIC couplings on different sublattices are oriented such that
they break chiral symmetry, and introduce a Haldane mass
$\Gamma(\nvec{k})/\omega$.
By formal comparison to the Haldane model, \eqref{eq:TCC Laplacian} amounts to inducing an
effective magnetic flux of $\phi =
\frac{\pi}{2}$~\cite{PhysRevLett.61.2015}. It is possible to
precisely control this fictitious magnetic flux in the TCC by a modification of the
impedance phase of the circuit elements in the INIC
(Fig.~\ref{fig:circuit_diagram}c). The Haldane mass induces a gapped
bulk admittance spectrum with chiral edge modes in the admittance gap
(Fig.~\ref{fig:band structure}a).

{\it Symmetries.}

The incorporation of resistances in a circuit environment, such as in the TCC, breaks time-reversal symmetry (TRS)~\cite{supp}.
From the viewpoint of thermodynamics, this is because
any resistive component experiences Joule heating, leading to increased entropy as well as
broken time-reversal symmetry. TRS translates into $J = -J^*$ in real space and $J(\nvec{k}) =
-J^*(-\nvec{k})$ in reciprocal space~\cite{supp}.  
We define circuit reciprocity as given by
$J^\top = J$ in real space and by
$J(\nvec{k}) = J^\top(-\nvec{k})$ in reciprocal space~\cite{supp}. 
By the use of operational amplifiers~\cite{marin2} as active
circuit elements, the INIC configuration acts as a charge
source or sink, causing an input or output
current from ground to the system. Our current feed from the INIC is arranged such that currents between two connected voltage nodes retain equal magnitude, but flow in opposite directions. This yields an antisymmetric contribution to $J_\text{TCC}$ and breaks reciprocity.

We name the TCC Hermitian if and only if $J_{\text{TCC}}$ is
anti-Hermitian, i.e., $J_{\text{TCC}} =-J_{\text{TCC}}^\dagger$, which leads to purely
imaginary admittance eigenvalues, and hence real
eigenfrequencies $\omega ({\bf k})$~\cite{supp}. The stationary time evolution of a given
TCC initial state is then conveniently expressed in terms of TCC
energy eigenmodes. $J_{\text{TCC}}$ is intimately connected to its
Hamiltonian formulation, as the eigenfrequencies $\omega(\nvec{k})$ are given by the
poles of the Greens function $G_{\text{TCC}}=J_{\text{TCC}}^{-1}$, and
hence relate to the roots of the admittance spectrum
$j_{\text{TCC}}(\omega ({\bf k}))=0$ (Fig. \ref{fig:band structure}b). 

\begin{figure*}[!t]
    \centering
    \includegraphics[width=\linewidth]{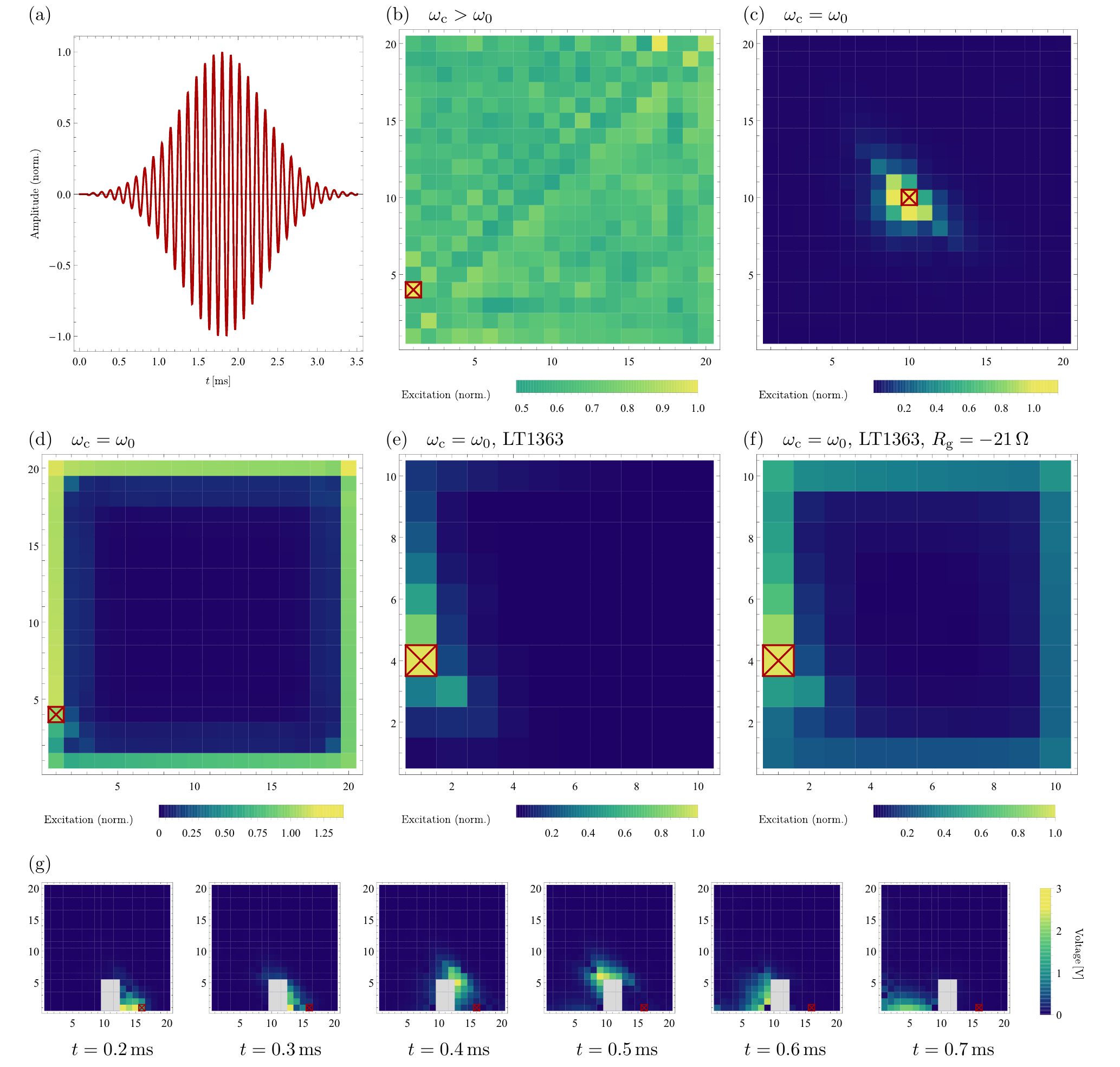}
    \caption{Simulated current pulse in a finite TCC. (a) An external
      current signal is fed in to the TCC. Its position in (b-g) is highlighted
      by a red crossed square, where all voltage profiles are
      normalized with respect to the input. (b-d) Integrated total voltage signal
      resolved at each unit cell and obtained from an \ltspice
      simulation for 20x20 unit cells with $A-B$ termination with
      ideal operational amplifiers and parasitic serial resistances of
      $R_{\text{$L_0$}}=\SI{1}{\milli \ohm}$ and
      $R_{\text{$C_0$}}=\SI{0}{\milli \ohm}$. Circuit components
      chosen as $C_0= \SI{10}{\micro \farad}$, $L_0= \SI{10}{\micro
        \henry }$, $R_0= \SI{10}{\ohm}$, $\Delta = \SI{0}{\farad}$,
      and $C_\text{g}= \SI{0}{\farad}$. (b) $(f_c, \Delta f_\text{exc})=(13.0,1.0)\,\text{kHz}$. The current
      spreads across the whole circuit. (c) $(f_c, \Delta
      f_\text{exc})=(9.2,0.3)\,\text{kHz}$. There is a localized circuit response upon bulk
      injection, as opposed to feeding into the chiral edge mode for a
      boundary injection (d). (e,f) $(f_c, \Delta f_\text{exc})=(290,10)\,\text{kHz}$. Integrated total voltage signal
      for $10 \times 10$ unit cells,
      $A-B$-termination, and realistic operational amplifiers
      LT1363. $C_0= \SI{0.1}{\micro \farad}$, $L_0= \SI{1}{\micro
        \henry }$, $R_0= \SI{30}{\ohm}$,
      $R_{\text{$L_0$}}=\SI{150}{\milli \ohm}$,
      $R_{\text{$C_0$}}=\SI{5}{\milli \ohm}$, $\Delta =
      \SI{0}{\farad}$, and $C_\text{g}= \SI{0}{\farad}$. In comparison
      to (e), (f) further implements INIC couplings to ground at
      each edge node, implying an effective negative resistance of $R_\text{g}= -\SI{21}{\ohm}$ for that connection. The chiral mode signal is significantly enhanced.
(g) Time-resolved voltage signal of the TCC with circuit parameters
identical to (d) and a defect area of size $(3 \times 5)$ unit cells
(grey region) by grounding the corresponding nodes.}
    \label{fig:Simulation}
\end{figure*}

{\it Topological phase diagram.} 
We define the Chern number for the lower admittance band $  C =
\frac{1}{2 \pi}  \oint\limits \dd^2 k \, \mathcal{B}(\nvec{k}),$ where $\mathcal{B}(\nvec{k})$ denotes the Berry
curvature~\cite{supp}. It is invariant under a change of the Bravais
vector gauge, whose only consequence is a distortion of the
Brillouin zone~\cite{helbig1}. As for the Haldane model, from gapping out the
two admittance Dirac cones due to finite Haldane or Semenoff mass, there
is a
topologically non-trivial regime with $C=1$ and a trivial regime with
$C=0$. We find
\begin{equation}
C = \frac{1}{2} \bigg[ \sign \left(\Delta + \frac{3 \, \sqrt{3}}{\omega R_0} \right) -\sign \left(\Delta - \frac{3 \, \sqrt{3}}{\omega R_0} \right) \bigg],
\end{equation}
which is nonzero if $  \omega R_0 < \frac{3 \, \sqrt{3} }{ \Delta}
$. In this case, placing oneself in the admittance or eigenfrequency
gap, and allowing for a boundary termination, one finds a chiral
mode located at the boundary (Fig.~\ref{fig:band structure}a and
Fig.~\ref{fig:band structure}b). The chiral voltage
boundary mode relates to $\omega(\nvec{k}) \neq \omega(-\nvec{k})$ associated with the breaking of circuit reciprocity~\cite{supp}.

{\it Circuit simulations.}
The TCC in principle allows for a detailed
characterization and calibration of the chiral voltage boundary mode. Even for a physical system as accessible
and tunable as
electric circuits, however, various types of imperfections have to be taken into account
to move from an ideal theoretical model to a realistic setting. This
includes circuit element variances, parasitic resistances, and other  
constraints on realistic operational amplifiers we use in the
INICs. The principal time scales and parametric dependencies of the
chiral edge mode can be
deduced from the clean TCC limit~\eqref{eq:band structure}, where we
set $\Delta=0$.  Within linear approximation of the edge mode in the frequency spectrum (Fig.~\ref{fig:band structure}b), the group velocities of the zigzag ($v_\text{zz}$) and bearded edge ($v_\text{bd}$) yield 
  $  v_\text{zz} = \frac{3 \sqrt{3}}{2 \pi} \frac{1}{R_0 \, C_0} = 2 \, v_\text{bd}$.
As seen from \eqref{eq:band structure}, the inverse
resistance $R_0^{-1}$ serves as a regulation parameter of the gap size.
Numerical analysis indicate, that for significantly small values of $R_0$ (such that $\omega R_0 C_0 $ 
is not significantly larger than unity), the edge mode localization length (understood in units of the internodal spacing) 
is tunable through the resistance $R_0$ in the INIC.

We create a finite TCC lattice, excite it with a Gaussian AC current signal
centered around $\omega_\text{c} =2 \pi \, f_\text{c}$ and a standard deviation $\Delta
\omega_\text{exc}=2\pi \, \Delta
f_\text{exc}$,  and perform \ltspice simulations on various
configurations (Fig.~\ref{fig:Simulation}). If, as in
Fig.~\ref{fig:Simulation}b, $\omega_\text{c}$ lies within the TCC
frequency band, 
the current spreads
across the whole circuit, even if it is injected at the boundary
(marked by a crossed square). If, however, $\omega_\text{c}$ lies
within the bulk gap and $\Delta
\omega_\text{exc}$ is sufficiently small, the circuit response is
crucially different depending on whether the current is injected in
the bulk (Fig.~\ref{fig:Simulation}c) or at the boundary
(Fig.~\ref{fig:Simulation}d). While it is localized for the former,
the signal propagates through the chiral edge mode along the boundary
for the latter. Note that due to parasitic effects introduced by the
serial resistances of inductors $R_{\text{$L_0$}}$ and capacitors
$R_{\text{$C_0$}}$, the voltage pulse in the circuit faces
dissipation caused by the shift of the resonance frequency spectrum along the
positive imaginary axis. The most relevant parasitic effects derive
from the inductor, introducing a timewise exponential decay constant $\tau =
\frac{2 \, L_0}{R_{\text{$L_0$}}}$ that damps the chiral voltage
signal. This realistic analysis is further refined in
Fig.~\ref{fig:Simulation}e and Fig.~\ref{fig:Simulation}f, where
we study no idealized, but explicit, publicly available
operational amplifier elements LT1363. As seen in Fig.~\ref{fig:Simulation}e, the
realistic setting experiences significant signal decay already across 10
unit cells. 
In Fig.~\ref{fig:Simulation}f, we illustrate one way to
calibrate the TCC towards a more stable edge signal by adding
INIC connections of effective negative resistance $R_\text{g}$ between the edge
nodes and ground, which effectively yields the insertion of a gain
parameter to the system. Through the adjustment of the resistive
components in the INICs connected to ground, one can conveniently
calibrate the given TCC realization closer towards its Hermitian point, which
enhances the Chern mode signal. This is only one of several ways to
improve the TCC Chern signal through an inherent TCC parameter
adjustment. Such a type of
gain implementation to compensate for dissipative loss has not been
accomplished in other Chern systems.
In Fig.~\ref{fig:Simulation}f, we implement a defect (grey) at the
boundary of the circuit, e.g. by grounding the corresponding voltage
nodes. In a time-resolved simulation, we observe a propagation of the
edge mode around this defect area. It demonstrates the topological
protection of the chiral edge mode, which roots in the existence of a
nonzero admittance Chern number as a bulk property.


{\it Conclusion.} 
We have introduced and analyzed the topolectrical Chern circuit as a topological
circuit array with active INIC circuit elements. A topological voltage Chern mode 
appears due to the non-reciprocity induced by the INICs, which at the
same time also serve as a convenient calibration tool to minimize the
dissipative loss of the edge mode.  This reaches an unprecedented level at which
a topological chiral edge mode is tunable and analyzable in all detail in an
accessible physical environment, and offers itself to further analysis
of topological circulator devices in general. 

\noindent


\section{Acknowledgments}
\begin{acknowledgments}
We thank S.~Imhof and A.~Stegmaier for helpful discussions.  The circuit simulations
have been performed by the use of \ltspice,
\url{http://www.linear.com/LTspice}. The work in W\"urzburg is
supported by the European Research Council (ERC) through
ERC-StG-Thomale-TOPOLECTRICS-336012 and by the German Research
Foundation (DFG) through DFG-SFB 1170, project B04.
\end{acknowledgments}



\clearpage

\appendix

\section{Appendix A: Negative impedance converter (INIC)}\label{app:nic}
\begin{figure}[!h]
    \centering
    \includegraphics[width=\linewidth]{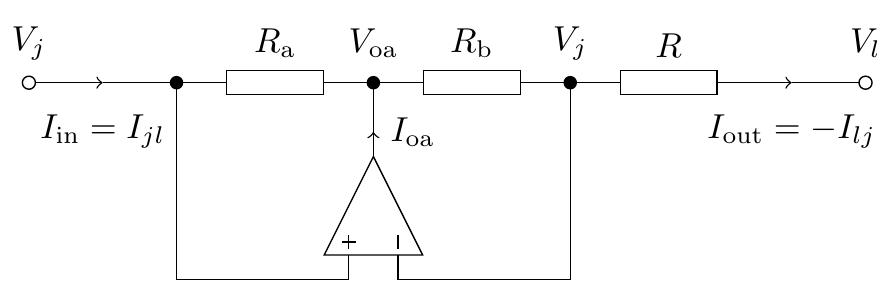}
    \caption{Circuit diagram of an operational amplifier in a negative impedance converter configuration with current inversion.}
    \label{fig:nic_diagram}
\end{figure}
For the realization of a negative impedance converter with current
inversion~\cite{Chen}, we use an operational amplifier (OpAmp)
configuration shown in Fig. \ref{fig:nic_diagram}. 
The current entering
the INIC from the left is given by $I_\text{in} = (V_j -
V_\text{oa})/R_\text{a}$, and the current leaving on the right by
$I_\text{out} = (V_j - V_l)/R$ when the OpAmp is operated in a negative feedback configuration. 
Assuming an infinite impedance of the OpAmp inputs forbids any current flowing into the OpAmp
and simplifies the output current to $I_\text{out} = (V_\text{oa} - V_j)/R_\text{b}$. Solving these equations for $I_\text{in}$ yields
\begin{subequations}
\begin{align}
    I_\text{in} \ \, &= -\frac{R_\text{b}}{R_\text{a}\cdot R} \left(V_j - V_l\right), \\
    I_\text{out} &= \frac{1}{R} \left(V_j - V_l\right).
\end{align}
\end{subequations}
Translating those results to the Laplacian form leads to the node voltage equation
\begin{align}
    \begin{pmatrix}
        I_{j l} \\ I_{l j}
    \end{pmatrix}
    = \frac{1}{R}
    \begin{pmatrix}
         -\nu & \nu \\-1 & 1 
    \end{pmatrix}
    \begin{pmatrix}
        V_j \\ V_l
    \end{pmatrix},
\end{align}
where $\nu = R_\text{b}/R_\text{a}$. The matrix is not symmetric, and
circuit reciprocity is thus broken. To obtain an anti-Hermitian
form as desired in the TCC, we require $\nu = 1$, \ie~
$R_\text{a} = R_\text{b}$. If this is not the case, the
eigenfrequencies of the system become complex, resulting in either a
damping or an instability. While the former is manageable, active
elements such as OpAmps always necessitate a careful setup such that
the active feed does not lead to a divergence, and hence breakdown of
the circuit. In mathematical terms, the eigenfrequencies should always
be adjusted such that their imaginary part is positive. 
In a real experiment, it hence needs to be ensured that the system operates
stable, i.e. accepting a certain overall loss in order to avoid
instabilities. Parasitic effects may in fact help to shift the circuit
towards the stable regime. Since electric circuits are so easily
tunable, and e.g. resistors can be implemented by individually
adjustable potentiometers, the interplay between robust stability and
low dissipation can be fine-tuned to the desired optimal Hermitian TCC
sweet spot.

\section{Appendix B: Hamiltonian formulation}
We define voltage and current
vectors by denoting the voltages measured at the nodes of a
circuit board against ground, and the input currents at the nodes by
$N$-component vectors $\nvec{V}$ and $\nvec{I}$, respectively. The
equations of motion of the circuit are given by
\begin{align}\label{eq:equation_of_motion2}
\frac{\dd}{\dd t} \nvec{I}(t) = C \frac{\dd^2}{\dd t^2} \nvec{V}(t) + \Sigma \frac{\dd}{\dd t} \nvec{V}(t) + L \,  \nvec{V}(t),
\end{align}
where capacitance $C$, conductance $\Sigma$, and inductance $L$ are the real-valued ($N \times
N$)-matrices forming the grounded circuit Laplacian~\cite{lee1} by
\begin{align} \label{eq:split_laplacian}
J(\omega) = \ii \omega \, C + \Sigma + \frac{1}{\ii \omega} \, L.
\end{align}
The homogeneous equations of motion ($\nvec{I} = 0$, where the circuit's
time evolution is solely determined by its eigenfrequencies) can be
re-written as $2N$ differential equations of first order 
\begin{align}\label{eq:Hamiltonian time evolution}
- \ii \, \frac{\dd}{\dd t} \psi(t) = H \, \psi(t),
\end{align}
where $\psi(t) :=(\dot{\nvec{V}}(t), \nvec{V}(t))^\top$, i.e., the voltages and their
first time derivatives are treated as independent variables.
This defines the $(2N\times 2N)$-Hamiltonian block matrix
\begin{align}\label{eq:Hamiltonian def}
H_{\text{TCC}} = \ii \begin{pmatrix}
C^{-1} \Sigma & C^{-1} L \\ -\mathbbm{1} & 0
\end{pmatrix}
\end{align}
where, for the TCC in reciprocal space as in Eq.~\ref{eq:TCC Laplacian}, 
\begin{subequations}
\begin{align}
C(\nvec{k}) &= \left(3 \, C_0 +C_\text{g}\right)\unity  +  C_0  \left(1 + \cos(k_x) + \cos(k_y)\right)\, \sigma_x 
\notag\\
&+ C_0 \left(\sin(k_x) + \sin(k_y)\right)\,  \sigma_y  +  \Delta \, \sigma_z, \\
\Sigma(\nvec{k}) &=\frac{2\, \im}{R_0} \left( \sin(k_x) - \sin(k_y) - \sin(k_x - k_y) \right)\,  \sigma_z, \\
L(\nvec{k}) &= \frac{1}{L_0} \, \mathbbm{1}.
\end{align}
\end{subequations}
The time evolution of a given eigenstate $\psi_{\alpha}(t)$ yields
$\psi_{\alpha}(t) = \psi_{\alpha} \, \e^{\im \omega_{\alpha} t}$
where the eigenvalues $\omega_{\alpha}$, $\alpha \in \{1, \dots, 2
N\}$ are the resonance frequencies of the system (Fig.~\ref{fig:band
structure}b), defined as the roots of the admittance eigenvalues
$j(\omega_{\alpha})=0$. 
To render the measurable voltage $\nvec{V}$ and its time derivative
$\dot{\nvec{V}}$ real, the eigenfrequencies of the Hamiltonian must
occur in pairs of $(\omega, -\omega^*)$ corresponding to
complex conjugated pairs of eigenstates $\psi$,
$\psi^*$. This enables us to label the eigenvalues by
$\omega_n^\pm$ with $\omega_n^- = - (\omega_n^+)^*$ and $n \in \{1,\cdots,N\}$. In Bloch form we label the
eigenfrequencies by their corresponding wavenumber $\nvec{k}$ and band
index $m$ as a composite index $n = (\nvec{k},m)$. This mapping from wave numbers to
frequencies as eigenvalues of the Hamiltonian defines the frequency spectrum $\omega(\nvec{k})$ in reciprocal space. 
Due to the intimate connection of the Laplacian and Hamiltonian spectrum, the emergence of a band gap in
the admittance band structure is accompanied by an analogous gap
opening in the frequency spectrum (Fig.~\ref{fig:band
structure}).
The eigenvectors of the Hamiltonian can be constructed out of the eigenvectors of the Laplacian according to 
\begin{align} \label{eq:H_eigenvectors}
\psi_n^+ = \begin{pmatrix} \ii \omega_n^+ \nvec{V}_n \\ \nvec{V}_n \end{pmatrix}
\quad
\text{and}
\quad
\psi_n^- = \begin{pmatrix} \ii \omega_n^- \nvec{V}_n^* \\ \nvec{V}_n^* \end{pmatrix}.
\end{align}
These are right eigenvectors. Note that the left eigenvectors of the non-Hermitian Hamiltonian may differ.

\section{Appendix C: Symmetries}

Written in the form of~\eqref{eq:split_laplacian}, it becomes clear
that the Laplacian is anti-Hermitian if and only if $C$ and $L$ are Hermitian
matrices while $\Sigma$ must be anti-Hermitian, which is fulfilled for
the ideal TCC Laplacian. 

Time reversal is defined as the parametric operation $t \rightarrow -t$. As a consequence, voltages transform even and currents odd under time reversal. Concluding from the time-reversed form of~\eqref{eq:equation_of_motion2}, we call a circuit  time reversal symmetric (TRS) if $\Sigma = 0$, which is equivalent to excluding any resistive component in the circuit, where energy can dissipate. This condition can be reformulated as allowing only for fully imaginary real space Laplacians, and is in accordance with the TRS condition $J^* = - J$ in real space defined in the main text, which can be checked by insertion into~\eqref{eq:split_laplacian}. Translated into reciprocal space, the TRS condition yields $J^*(\nvec{k}) = - J(-\nvec{k})$.

As a visualization of circuit reciprocity, consider two voltage nodes of the circuit with labels $j$ and $l$ connected by a passive circuit element. The current flowing out of node $j$ must be the current entering at node $l$. In other words, the current running from $j$ to $l$, $I_{j l}$, is the negative of the current running from $l$ to $j$, \ie~$I_{j l} = -I_{l j}$. This behaviour is called circuit reciprocity. In the Laplacian formalism, circuit reciprocity is given by $J_{j l} = J_{l j}$ or $J^\top = J$ in real space, and by $J(\nvec{k}) = J^\top(-\nvec{k})$ in Bloch form. 
In the present implementation, we use operational amplifiers as active
circuit elements in an INIC configuration to tune the system to the
point where $I_{j l} = I_{l j}$ for the INIC connection. Other
implementations of broken circuit reciprocity use magnetic fields or
sophisticated transistor arrays such as in passive circulators
developed for electrical communication engineering. Based on the fact
that the resonance frequencies for
$j(\omega,\nvec{k})=j(\omega,-\nvec{k})=0$ are identical to the
eigenvalues of the Hamiltonian, we obtain  $\omega(\nvec{k}) =
\omega(-\nvec{k})$ for the frequency spectrum. 
The voltage eigenmodes of a periodic system are constructed out of Bloch waves in reciprocal space and retrieve the form of plane waves labeled by the wave number $\nvec{k}$,
\begin{align}
    \nvec{V}_{\nvec{k},m}(\nvec{x},t) = \nvec{V}_{\nvec{k},m} \, \e^{\ii \nvec{k} \cdot \nvec{x}} \, \e^{\ii \omega_m(\nvec{k}) t}.
\end{align}
For the following considerations, we assume a Hermitian circuit system. If it is additionally TRS and reciprocal, the eigenmodes for inverse $\nvec{k}$ are characterized by
\begin{align}
    \nvec{V}_{-\nvec{k},m}(\nvec{x},t) = \nvec{V}_{\nvec{k},m} \, \e^{-\ii \nvec{k} \cdot \nvec{x}} \, \e^{\ii \omega_m(\nvec{k}) t}
\end{align}
by using $\nvec{V}_{-\nvec{k},m} = \nvec{V}_{\nvec{k},m}$ and
$\omega_m(\nvec{k}) = \omega_m(-\nvec{k})$. The
eigenmodes to $-\nvec{k}$ travel in the opposite direction than the
ones corresponding to $\nvec{k}$, and ultimately combine to standing
waves that experience no propagation in any direction in a
steady-state solution of the system. 
A propagation can only be induced by breaking the symmetry $\omega(-\nvec{k}) = \omega(\nvec{k})$, which is solely possible by a combined breaking of time reversal symmetry and reciprocity. As detailed in the main text, the INIC configuration implemented as the next-nearest neighbor hopping element breaks time reversal symmetry and reciprocity simultaneously, inducing 
unidirectional propagation in one direction along the boundary of the
TCC in the topological regime. 
\begin{figure}[!h]
	\centering
	\includegraphics[width=\linewidth]{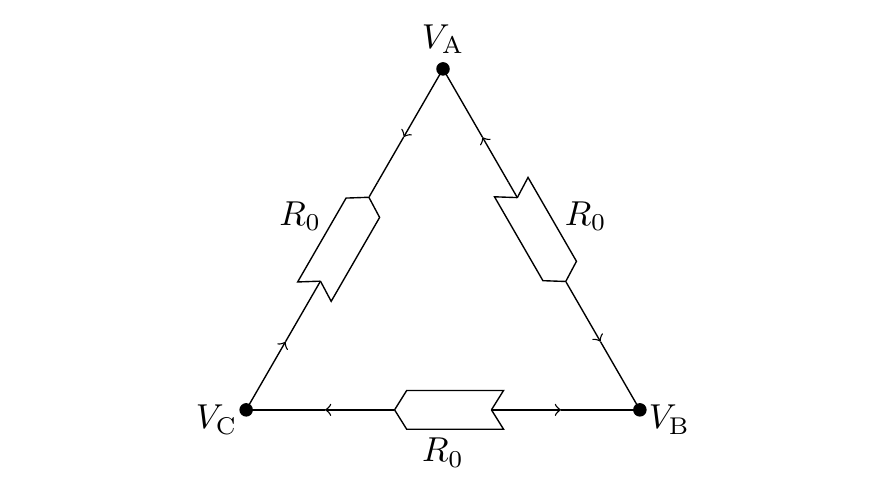}
	\caption{Triangular oriented INIC
        arrangement with resistances $R_0$. This network acts as a voltage circulator which breaks
        chiral symmetry and reciprocity. In the present scheme, we choose $V_\text{A} > V_\text{B} > V_\text{C}$ and mark the direction of the emerging currents in this configuration by arrows. }
	\label{fig:INIC triangle}
\end{figure}
The states associated with the midgap admittance and frequency bands
in Fig.~\ref{fig:band structure}a and Fig.~\ref{fig:band structure}b of the main text
are localized at opposing edges of the circuit due to the breaking of
chiral symmetry $\mathcal{C}: \{\sigma_z, J_{\nvec{k}} \} = 0$ through the mass term $ \ii \left(\Delta  +
  \Gamma(\nvec{k})/\omega \right)\, \sigma_z$  defined in
the main text. In the TCC model, $\Gamma(\nvec{k})$ establishes the
nontrivial topology through next-nearest neighbor INIC connections in
a three node arrangement as seen in Fig.~\ref{fig:circuit_diagram}b
and in a more highlighted form in Fig.~\ref{fig:INIC triangle}. The
associated breaking of reciprocity is caused by a circular motion of
voltage in the bulk of the TCC. To physically explain this behaviour,
consider three nodes $A,B,C$ in Fig.~\ref{fig:INIC triangle} with
corresponding voltages $V_\text{A} > V_\text{B} > V_\text{C}$. In this
setting, a current flows out of INIC $AB$ and $BC$, but into $CA$ from
both sides, respectively as depicted in Fig.~\ref{fig:INIC triangle}. This way, the current flows from both sides into node $B$, and $V_B$ surpasses the voltage $V_A$, such that INIC $AC$ now acts as a current drain and current flows out of node $A$ in both directions. Consequently, $V_A$ will fall below $V_C$, and a new configuration $V_B > V_C > V_A$ is reached. The voltage has thereby traveled in a clockwise direction as indicated by the orientation of the INICs in Fig.~\ref{fig:INIC triangle}. This process will continuously repeat such that a circular motion of voltage in this triangular INIC connection is established, thereby breaking chiral symmetry and reciprocity. 
The corresponding eigenmodes attain a chiral character, and while applying open boundary conditions, it becomes possible to initiate a voltage wave packet at the boundary of the TCC, which travels along the edge of the circuit in one direction.

\section{Appendix D: Hermitian TCC limit}
\paragraph{Lemma.} 
Let a circuit network consist of $N$ nodes described by a circuit Laplacian $J$ in real space, which can be divided into contributions of different circuit elements according to~\eqref{eq:split_laplacian}. Moreover, let all capacitances and inductances in the Laplacian be positive. If the Laplacian matrix is anti-Hermitian, $J = - J^\dagger$ for all real frequencies $\omega$, then the eigenfrequencies of the corresponding Hamiltonian \eqref{eq:Hamiltonian def} are real.

\paragraph{Proof.}
If $J = -J^\dagger, \; \forall \, \omega \in \mathbbm{R}$, it means that $C = C^\dagger$, $L = L^\dagger$ and $\ii \Sigma = (\ii \Sigma)^\dagger$. Consider the quadratic forms 
\begin{subequations}
\begin{align}
    c(\nvec{V}) = \nvec{V}^\dagger \, C \, \nvec{V}, \\
    \sigma(\nvec{V}) = \nvec{V}^\dagger \, \ii\Sigma \, \nvec{V}, \\
    l(\nvec{V}) = \nvec{V}^\dagger \, L \, \nvec{V} ,
\end{align}
\end{subequations}
which are real for all complex $N$-vectors $\nvec{V}$, since the representing matrices are Hermitian. 
Assuming only positive capacitances and inductances in the system, the structure of the Laplacian matrices dictates that $C$ and $L$ are weakly diagonally dominant.
Moreover, all entries of the matrices are real, and the diagonal elements are 
non-negative. Therefore, the matrices are positive semi-definite and $c(\nvec{V}), l(\nvec{V}) \geq 0, \; \forall \, \nvec{V} \, \in \, \mathbbm{C}^N$.

Now assume that $\psi_n = (\ii \, \omega_n V_n, V_n)^\top$ is an eigenvector of the Hamiltonian to eigenvalue $\omega_n$. Then the equation
\begin{align}
    \ii \omega_n \, c(\nvec{V}_n) - \ii \, \sigma(\nvec{V}_n) + \frac{1}{\ii \omega_n} \, l(\nvec{V}_n) = 0
\end{align}
must hold. For the non-trivial cases $\omega_n \neq 0$ and $c(\nvec{V}_n) \neq 0$, we can solve the quadratic equation for the eigenfrequency by completing the square. We obtain
\begin{align}
    \omega_n = \frac{1}{c(\nvec{V}_n)} \left( \sigma(\nvec{V}_n)  \pm \sqrt{\sigma^2(\nvec{V}_n) + 4 \, c(\nvec{V}_n) \, l(\nvec{V}_n)} \right) 
\end{align}
which is real, because all components are real-valued and $c(\nvec{V}_n)$, $l(\nvec{V}_n)$ are nonnegative. $\square$

\section{Appendix E: Berry phase in Hamiltonian and Laplacian form}
The Chern number as a topological invariant is defined for the
Hamiltonian as the generator of time translation. In the following, we show that for a time-independent, Hermitian system, $J = - J^\dagger$, the Berry phase and the Chern number, 
which determine the topological phase of the system, can be identically 
computed using the Laplacian eigenvectors. Consider the Berry connection $\mathcal{A}_n(\nvec{k})$ defined as
\begin{align}
    \mathcal{A}_m(\nvec{k}) = \ii \, \psi_m^\dagger(\nvec{k}) \, \partial_\nvec{k} \, \psi_m(\nvec{k}).
\end{align}
Here, the $\psi_m$s are the right eigenvectors of the Hamiltonian. Although the system is Hermitian ($J^\dagger = - J)$, 
the left eigenvectors of the Hamiltonian may differ from the right eigenvectors. The latter can be expressed using the eigenvectors $\nvec{V}_m$ of the Laplacian, 
\begin{align}
    \psi_m(\nvec{k}) 
    = 
    \frac{1}{N_m(\nvec{k})} 
    \begin{pmatrix} 
        \ii \, \omega_m(\nvec{k}) \nvec{V}_m(\nvec{k}) \\ 
        \nvec{V}_m(\nvec{k}) 
    \end{pmatrix},
\end{align}
where the $V_m(\nvec{k})$ are assumed to be normalized, such that $N_m(\nvec{k}) = \sqrt{\omega_m^2(\nvec{k}) + 1}$ is the normalization of the Hamiltonian eigenvectors. 
Computing the derivative of the eigenvector while omitting the band indices that are irrelevant for the computation yields 
\begin{align}
    \partial_\nvec{k} \psi 
    = 
    \frac{1}{N} \left[ 
        \begin{pmatrix}
            \ii \, (\partial_\nvec{k} \omega)\, \nvec{V} \\ 0
        \end{pmatrix}
        +
        \begin{pmatrix} 
            \ii \omega \, \partial_\nvec{k} \nvec{V} \\
            \partial_\nvec{k} \nvec{V}  
        \end{pmatrix} 
        - \psi \, \partial_\nvec{k} N 
    \right].
\end{align}
The projection of this result on $\psi^\dagger$, exploiting the normalization of the Laplacian eigenvectors, reads 
\begin{align}
    \psi^\dagger \partial_\nvec{k} \psi
    = 
    \frac{1}{N^2} \left[ 
        \omega \, \partial_\nvec{k} \omega 
        + N^2 \, \nvec{V}^\dagger \partial_\nvec{k} \nvec{V} 
        - N \, \partial_\nvec{k} N
    \right].
\end{align}
Since $N \, \partial_\nvec{k} N = \omega \, (\partial_\nvec{k} \omega)$, the first and last term in the bracket cancel, and we eventually find 
\begin{align}
    \mathcal{A}_m(\nvec{k}) = \ii \, \nvec{V}_m^\dagger(\nvec{k}) \, \partial_\nvec{k} \, \nvec{V}_m(\nvec{k}).
\end{align}
That means that the Berry connection and therefore the Berry curvature and Chern number for the (right) Hamiltonian and Laplacian eigenvectors coincide. Note that the left eigenvectors or right and left eigenvectors of $H$ combined may give alternative Berry connections and curvatures. However, the associated Chern numbers are equal~\cite{PhysRevLett.120.146402}. In this regard, the admittance and frequency band structure of a Hermitian system is topologically equivalent, and we can define the admittance Chern number as the global topological invariant of a circuit system.

\section{Appendix F: Low-admittance TCC expansion}
The TCC Laplacian in its two-band Bloch from in \eqref{eq:TCC
  Laplacian} of the main text can be recast in terms of the Pauli
matrices $\sigma_x$, $\sigma_y$ and $\sigma_z$ as
$J_{\text{TCC}}(k_x,k_y) =\ii \left( d_0 + \vec{d}(\nvec{k}) \cdot
  \vec{\sigma} \right)$, 
\begin{subequations}
\begin{align}
    d_0 &= \omega  \left( 3 \, C_0 + C_\text{g} -\frac{1}{\omega^2 \, L_0}\right)\\
    d_x &= -\omega \, C_0 \left(1 + \cos(k_x) + \cos(k_y)\right)\\
    d_y &= -\omega \, C_0 \left(\sin(k_x) + \sin(k_y)\right)\\
    d_z &= \omega  \left( \Delta + \frac{2}{\omega \, R_0} \left( \sin(k_x) - \sin(k_y) - \sin(k_x - k_y) \right)\right).
\end{align}
\end{subequations}
It features an inversion symmetry breaking Semenoff mass $\ii  \omega
\, \Delta$ and a reciprocity and time reversal symmetry breaking
Haldane mass $\ii \, \Gamma(\nvec{k})/\omega$, which both open a band gap due to their appearance in $d_z(\nvec{k})$.  We can compute the effective low admittance theory at the Dirac cones at $\nvec{K}_+ = (2\pi/3, 4\pi/3)$ and $\nvec{K}_- = (4\pi/3, 2\pi/3)$ by expanding $\nvec{k} = \nvec{K}_{\pm} + \nvec{q}$ up to first order in $\nvec{q}$ as
\begin{align*}
    J_{\text{TCC,eff}}^{(\pm)}(\nvec{q}) = &\ii \omega \Big[ \left( 3 C_0 +C_g -\frac{1}{\omega^2 \, L_0}\right) \mathbbm{1} \\
    &\mp C_0\frac{\sqrt{3}}{2} \, (q_y-q_x) \sigma_x + C_0 \, \frac{1}{2} \, (q_x+q_y) \, \sigma_y \\
    &+\left(\Delta \pm \frac{3 \, \sqrt{3}}{\omega R_0}  \right) \sigma_z \Big]+\mathcal{O}(\nvec{q}^2),
\end{align*}
where the $\pm$-signs represent the positive and negative chirality of
the Dirac cones with admittance gapping mass terms of
$m_{\pm}=\Delta\pm \frac{3 \, \sqrt{3}}{\omega R_0} $,
respectively. To establish more of a form invariance to the
conventional Dirac form, we implement a linear transformation in
$(q_x,q_y)$ via $    \frac{\sqrt{3}}{2} (q_y-q_x) \rightarrow p_y,
    \frac{1}{2} (q_x+q_y)  \rightarrow p_x$,
which being a gauge
transformation for the circuit graph manifests as a distortion of the Brillouin zone. Neglecting the term proportional to the unit matrix gives the transformed effective low admittance TCC Laplacian 
\begin{widetext}
\begin{align}
    \frac{1}{\ii \omega} \ J_{\text{TCC, eff}}^{(\pm)}\,(\nvec{p}) = \mp \, C_0 \, p_y \,  \sigma_x + C_0 \, p_x \, \sigma_y +\left(\Delta \pm \frac{3 \, \sqrt{3}}{\omega R_0}  \right) \sigma_z + \mathcal{O}(\nvec{p}^2).
\end{align} 
\end{widetext}
The sign of the Haldane mass $\Gamma(\nvec{K}_{\pm})/\omega = \pm \frac{3 \,
  \sqrt{3}}{\omega R_0}$ relates to the chirality of the Dirac cones,
effectively yielding mass terms with opposing signs at the Dirac
cones, while the Semenoff mass stays invariant throughout the whole
BZ. The Haldane mass therefore induces a Berry flux monopole to which
each Dirac cone contributes one half integer Berry charge. The mapping $T^2 \rightarrow S^2 \, : \, (k_x,k_y) \mapsto \hat{\nvec{d}} = \vec{d}/\left|d\right|$ from the torus to the unit sphere establishes the Chern number as the winding number of the vector $\hat{\nvec{d}}$ as a covering of the unit sphere. In the TCC model, this winding number is $1/2$ for each Dirac cone due to the half covering of the unit sphere by the mappings of $\hat{\nvec{d}}$ for each cone. 
The Chern number of the admittance band structure defined in a gauge-invariant scheme resorts to the introduction of the Berry curvature
\begin{align}\label{eq: Berry curvature}
    \mathcal{B}(\nvec{k}) = \frac{1}{2} \,  \hat{\nvec{d}} \left ( \frac{\partial \hat{\nvec{d}}}{\partial k_x} \times \frac{\partial \hat{\nvec{d}}}{\partial k_y} \right) .
\end{align}
We now illustrate that in a reciprocal or time reversal
symmetric system, the Chern number vanishes due to the antisymmetry of
the Berry curvature in reciprocal space. For that, recall the condition $J^\top(-\nvec{k}) = J(\nvec{k})$ for the circuit Laplacian of a reciprocal system. This leads to individual constraints on the components of the $\nvec{d}$-vector 
\begin{subequations}
\begin{align}
    &d_0 (-\nvec{k}) =    d_0 (\nvec{k}),\\
    &d_x (-\nvec{k}) =    d_x (\nvec{k}), \label{eq:rez constraint d_x}  \\
    &d_y (-\nvec{k}) =   - d_y (\nvec{k}), \\
    &d_z (-\nvec{k}) =   d_z (\nvec{k}).  \label{eq:rez constraint d_z} 
\end{align}
\end{subequations}
We can then use these conditions to compute the Berry curvature \eqref{eq: Berry curvature} of a reciprocal system for inverse wave vectors
\begin{align}
    \mathcal{B}(-\nvec{k}) &= \frac{1}{2} \, \hat{\nvec{d}}(-\nvec{k}) \left ( \frac{\partial  \hat{\nvec{d}}(-\nvec{k})}{\partial k_x} \times \frac{\partial  \hat{\nvec{d}}(-\nvec{k})}{\partial k_y} \right) \nonumber \\
    &= - \mathcal{B}(\nvec{k}) \label{eq: Berry curvature odd in k}
\end{align}
by insertion of~\eqref{eq:rez constraint d_x} - \eqref{eq:rez constraint d_z}. In a time-reversal symmetric system, exploiting the condition $-J^*(-\nvec{k}) = J(\nvec{k})$,
\begin{subequations}
\begin{align}
    &d_0^* (-\nvec{k}) =    d_0 (\nvec{k}),\\
    &d_x^* (-\nvec{k}) =    d_x (\nvec{k}), \label{eq:trs constraint d_x}  \\
    &d_y^* (-\nvec{k}) =   - d_y (\nvec{k}), \\
    &d_z^* (-\nvec{k}) =   d_z (\nvec{k}).  \label{eq:trs constraint d_z} 
\end{align}
\end{subequations}
For Hermitian systems, the conditions \eqref{eq:trs constraint d_x}
-\eqref{eq:trs constraint d_z} equal those for reciprocal systems,
because the $\nvec{d}$-vector must be real. In such a case,~\eqref{eq:
  Berry curvature odd in k} is trivially fulfilled. The TCC model is a Hermitian model by construction, which means that while omitting next-nearest neighbor hopping terms, the Chern number must vanish. 
Only by breaking time-reversal symmetry and reciprocity through the
INIC next-nearest neighbor connection can the Chern number be
nonzero. From a topological point of view, this establishes the
combined breaking of reciprocity and time reversal symmetry in a
circuit context as the analogue of time reversal symmetry breaking in
a quantum electronic system for which the Haldane model was formulated~\cite{PhysRevLett.61.2015}.

In Appendix D, we have shown that the Berry phase and, consequently,
the Chern number is identical in the Hamiltonian and Laplacian
formalism. We can therefore establish the admittance band Chern number
as a topological invariant of the circuit dynamics. In our TCC model,
we set the phase originating from the effective magnetic
flux~\cite{PhysRevLett.61.2015} in the next-nearest neighbor hopping
term to $\phi = \frac{\pi}{2}$ , while in full generality, an arbitrary
phase $\phi$ is possible by incorporating the desired complex
impedance in the INIC configuration. The value of $\frac{\pi}{2}$ is
the favoured configuration in our context, as it features the largest bandgap. By terminating the circuit in an arbitrary direction, we expect to acquire edge modes that are exponentially localized at the boundary between a topologically nontrivial and trivial regime.

\begin{figure}[!h]
	\centering
	\includegraphics[width=\linewidth]{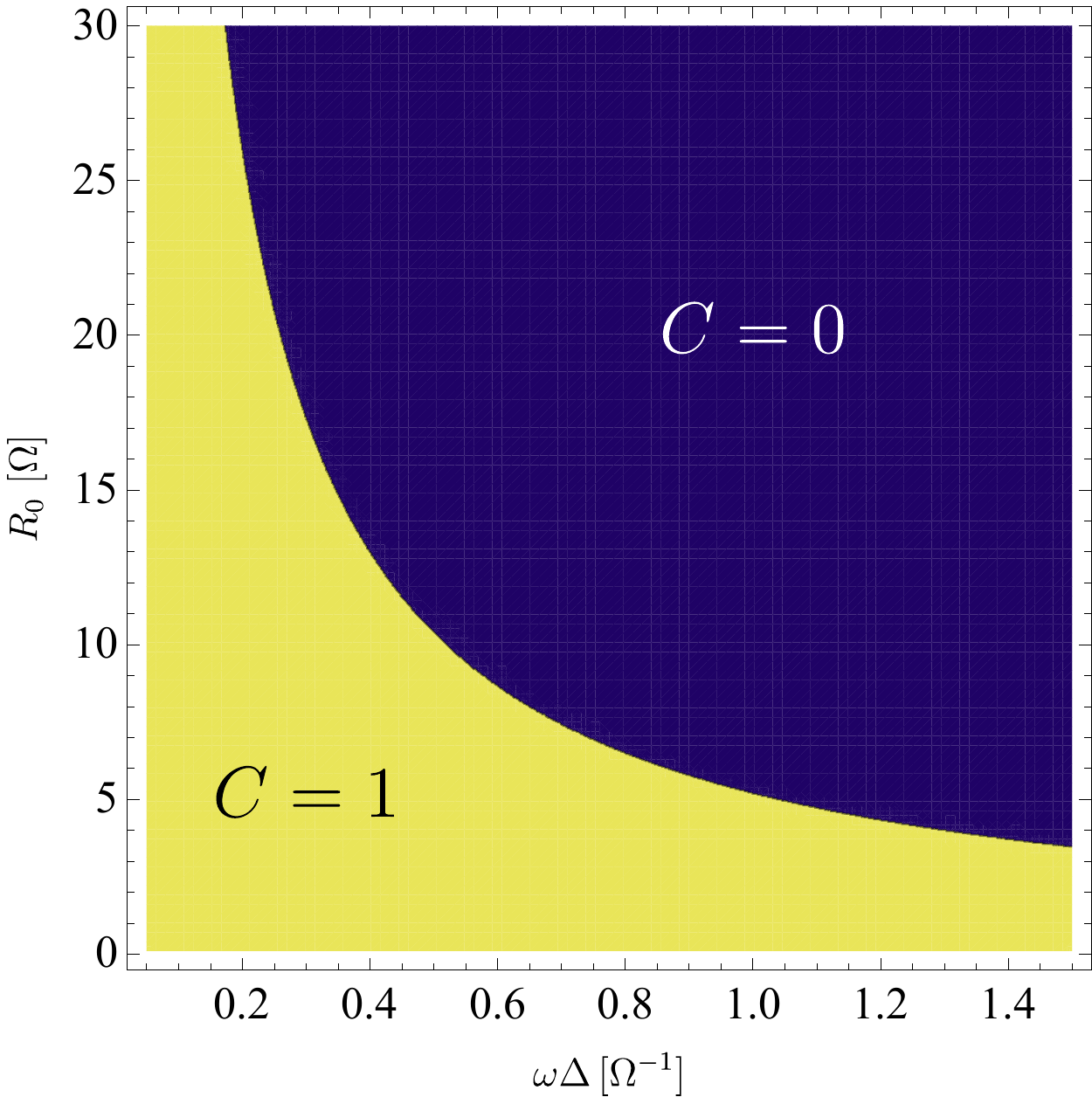}
	\caption{Topological phase diagram of the TCC model as a function of the combination of frequency and capacitive grounding term $\omega \,\Delta$ and the INIC resistance $ R_0$, both restricted to positive values only. The topological regime is marked by the Chern number $C=1$, whereas the topologically trivial phase has vanishing Chern number $C=0$.}
	\label{fig:phase_diagram}
\end{figure}

The Chern number for the lower band of the TCC model amounts to the combination of the Chern numbers caused by the half-charged Berry flux monopoles at each Dirac cone 
\begin{align*}
C &= \frac{1}{2} \left( \sign \left(m_+ \right) -\sign \left(m_- \right) \right)\\
&= \frac{1}{2} \left( \sign \left(\Delta + \frac{3 \, \sqrt{3}}{\omega R_0} \right) -\sign \left(\Delta - \frac{3 \, \sqrt{3}}{\omega R_0} \right) \right),
\end{align*}
which is nonzero if $\sign \left(\Delta - \frac{3 \, \sqrt{3}}{\omega R_0}  \right) < 0$ as mentioned in the main text. In Fig.~\ref{fig:phase_diagram}, we show the phase diagram of the TCC model as a function of the capacitive grounding term $\Delta$ and the combination of frequency and INIC resistance $\omega \, R_0$, which we both restrict to positive values. The topological regime is marked by the Chern number $C=1$, whereas the topologically trivial phase has vanishing Chern number. As expected, in the limit of $\Delta \rightarrow 0$, we are always in the topological regime, no matter how we choose $\omega \, R_0$. For larger $\Delta$, the parameter $\omega \, R_0$ needs to be reduced in an inversely proportional fashion to $\Delta$ in order to stay in the topological regime. The line of phase transition is determined by the condition $ R_0 = \frac{3 \sqrt{3}}{\omega \, \Delta}$.

\clearpage

\end{document}